\documentclass{article}
\input amssym.def
\input amssym.tex
\usepackage[a4paper]{geometry}
\usepackage{graphicx, color}
\usepackage{multicol}
\def\ben{\begin{equation}}
\def\een{\end{equation}}
\def\half{\frac{1}{2}}
\def\bea{\begin{eqnarray}}
\def\eea{\end{eqnarray}}

\def\bx{{\bf x}}

\def\bn{{\bf n}}

\def\p{\partial}
\def\mathbb{\Bbb}

\def\ben{\begin{equation}}
\def\een{\end{equation}}
\def\half{{1 \over 2}}
\def\bea{\begin{eqnarray}}
\def\eea{\end{eqnarray}}

\def\bn{{\bf n}}
\def \p{\partial}
\def\p{\partial}
 
\def\bom{{\mbox{\boldmath $ \omega $ }}}

\def\x{{\bf x}}\def\B{{\bf B}}

\def\n{{\bf n}}

\makeatletter

\newenvironment{figurehere}
  {\def\@captype{figure}}
  {}
\makeatother

\newcount\hour \newcount\minute
\hour=\time  \divide \hour by 60
\minute=\time
\loop \ifnum \minute > 59 \advance \minute by -60 \repeat
\def\nowtwelve{\ifnum \hour<13 \number\hour:
                      \ifnum \minute<10 0\fi
                      \number\minute
                      \ifnum \hour<12 \ A.M.\else \ P.M.\fi
         \else \advance \hour by -12 \number\hour:
                      \ifnum \minute<10 0\fi
                      \number\minute \ P.M.\fi}
\def\nowtwentyfour{\ifnum \hour<10 0\fi
                \number\hour:
                \ifnum \minute<10 0\fi
                \number\minute}

\title{The geometry of sound rays in a wind}
\author{G. W. Gibbons$^1$ and  C.M. Warnick$^{1, 2}$ 
\\
\\ \small{1. D.A.M.T.P., Cambridge, Wilberforce Road, Cambridge CB3 0WA,
  U.K.}
\\ \small{2. Queens' College, Cambridge, CB3 9ET, U.K.} \\}

\begin{document}
\maketitle {\let\thefootnote\relax\footnotetext{{\em Emails}:
  g.w.gibbons@damtp.cam.ac.uk, c.m.warnick@damtp.cam.ac.uk \\ \mbox{} \hspace{.45cm}\emph{Pre-print no.} DAMTP-2011-11}}
\begin{abstract}
We survey the close relationship between sound and light rays and
geometry. In the case where the medium is at rest, the
geometry is the classical geometry of Riemann. In the case where the
medium is moving, the more general geometry known as Finsler geometry
is needed. We develop these geometries ab initio, with examples, and in
particular show how sound rays in a stratified atmosphere with a wind
can be mapped to a problem of circles and straight lines.
\end{abstract} 


\begin{multicols}{2}
\section{Introduction: Propagation of Waves in Moving Media}

\begin{figure*}
\begin{minipage}[t]{0.48\linewidth}
\centering 
\begin{picture}(0,0)%
\put(-95, 30){\includegraphics[width=2.5in]{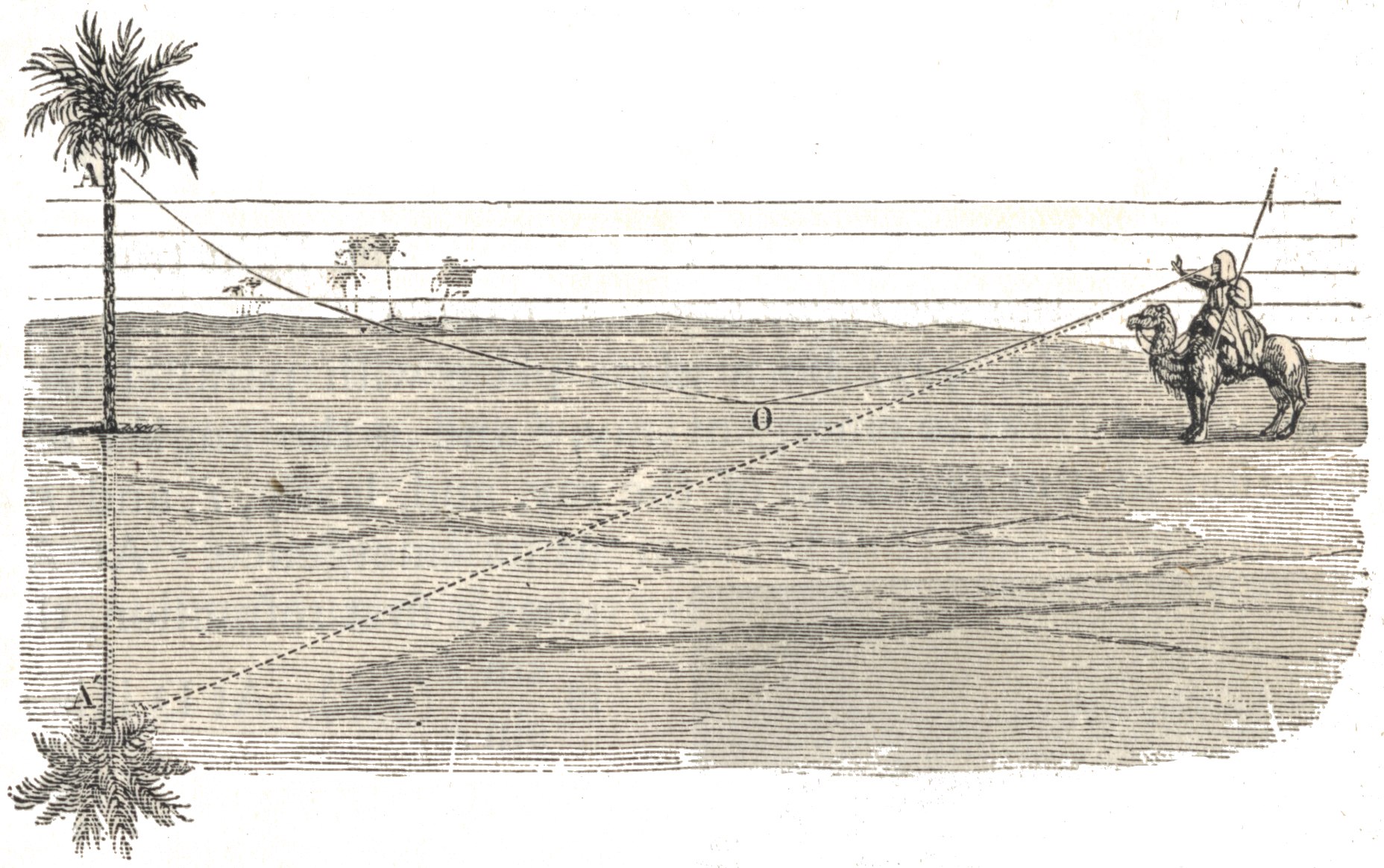}}%
\end{picture}%
\caption{A  19$^{th}$ century woodcut showing a mirage in the
  desert. \emph{\small Source:
  `\'El\'ements de Physique', Ganot} \label{mirage}}
\end{minipage}
\hspace{.3cm}
\begin{minipage}[t]{0.48\linewidth}
\centering {\includegraphics[width=2.5in]{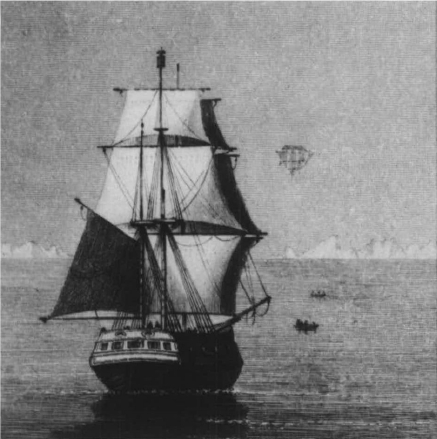}}
\caption{Etching of a sketch by the explorer William Scoresby, showing an arctic mirage. \emph{\small Source: `The life of William
Scoresby', Scoresby-Jackson}\label{fata}}
\end{minipage}
\end{figure*}

Almost everyone must be familiar, by report at least, 
with optical mirages. In hot regions such as the desert,
when the speed of light is greater at low altitudes than at higher
altitudes, distant palm trees can appear inverted as if reflected off a cool pool
of  water  in a nearby oasis, see Figure \ref{mirage}. It is less well known
that in cold regions such as the arctic, where the opposite
conditions prevail, distant ships can appear 
inverted in the sky, the light from them having been bent over an
intervening iceberg, see Figure \ref{fata}.
More mundanely, motorway  travellers on hot dry summer days often have
the disconcerting impression that there are sheets of water lying on the road
some distance  ahead. The explanation of phenomena like this
is easily understood  using the concept of light rays subject to  
Snell's Law for a stratified medium. Alternatively we may apply  
Huygens's  wave theory  according to which, in passing
over the iceberg, the higher parts of the wave front move faster 
than the lower parts causing it to bend downwards.

{\it Mutatis Mutandis} similar phenomenon can apply to sound waves
whose speed increases with increasing absolute temperature above zero,
$T$, as $T^\half$.  During a warm sunny day
therefore, when the temperature is typically hotter near  the ground
due to the sun's rays, the sound waves should be bent upwards.  On a
clear night, when the temperature near  the ground  drops  
rapidly by  radiative cooling, sound waves should be bent downwards
allowing distant sources, such as cars on a motorway,  to be heard        
much more clearly than during the day. One of us, living as he does a 
kilometre or so away from the A14, which is, judged by the amount 
of traffic it carries  effectively a motorway, had until recently always
supposed that it was this temperature effect that was responsible for
the din experienced  at times when contemplating the night sky 
in his garden. However, the temperature  effect should be isotropic,
in other words it should affect the noise of cars coming from all directions
and including those on less busy but closer city roads.  The greater  volume
and speed of traffic on the A14 alone should not be so overwhelming.

The obvious directional influence is that of the wind.
However, the speed of sound $v_s \approx 1250 $ km  per hour
is  so much greater than typical wind speeds $v \approx 30 $ km per hour
that simple  convection of sound waves by the wind cannot be
responsible for any significant directional effect. As
pointed out to one of us by a colleague, Hugh Hunt, who made
a similar point in the New Scientist of 15th April 2009  in response 
to readers' queries, it is not the wind velocity, but its gradient, that is its {\sl variation
with height}, called {\it wind shear} or  more  technically
{\it vorticity}   which is important. As is clear from observing
clouds, wind speeds, while remaining roughly horizontal and much slower than the speed of sound,
increase quite sharply with height, while remaining roughly horizontal. 
Thus it is not just the increased velocity $|{\bf v}|$ 
that matters, but its gradient or vorticity $\bom  = {\rm curl}\, {\bf
  v}$.

\begin{figure*}
\centering
\includegraphics[width=.9\linewidth]{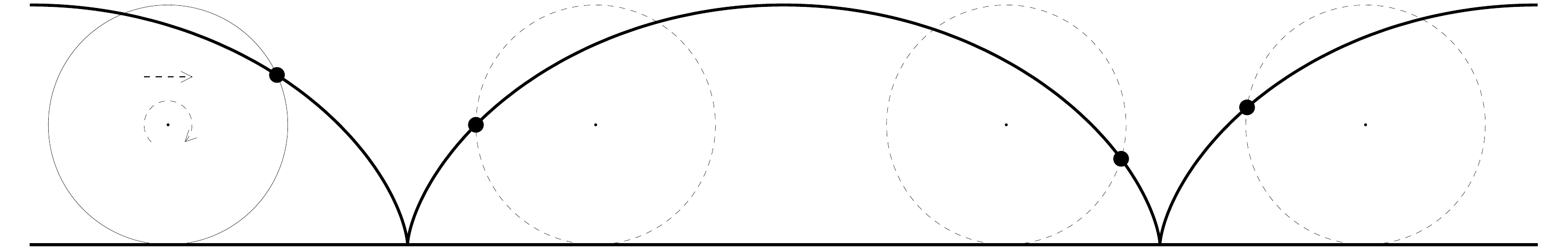}
\caption{The cycloid, as traced out by a point on the circumference of
  a rolling circle \label{cyclfig}}
\end{figure*}

Of course, examination of the literature
shows that this is not a new observation
and many papers and textbooks contain a simple qualitive discussion.
The earliest we have discovered  dates from 1857 \cite{Stokes}  and 
is due George Gabriel Stokes (1819-1903) who was appointed to the
Lucasian Chair of Mathematics in Cambridge in 1849 and held it until
his death 54 years later.  His explanation, elaborated on by Osborne
Reynolds (1842-1912) in 1874 \cite{Reynolds},  
followed very closely Huygens's explanation
for refraction by a gradient in the refractive index. If the wind direction
is towards us, and the wind speed increases with height, the higher parts of
the wave front  move faster than  the lower parts and the wave  
front is  bent over towards us. If on the other hand the 
wind direction is in the opposite direction,  the wave  front
will be bent upwards and hence away from us.  The net effect is that
if the vorticity $\bom$ is non zero, 
the sound rays  are deflected  in an analogous  way
to the deflection of a charged  particle of mass $m$ and charge $e$ 
by the Lorentz Force due to  magnetic field $\B$. In fact this is not just an analogy
but, as we shall see  shortly, a precise correspondence for low wind speeds: 
\ben
\bom \equiv - \frac{e}{m} \B \,. \label{vort}
\een       
 Note that since it is the velocity gradient that matters,
it is not the direction of the wind at the ground or 
at a height above , but their difference which is important.
In practice however, the speed  of the wind is 
almost always much slower near the ground than above,
and so usually it is the velocity at higher altitudes which matters.  
Thus, in principle there are two effects acting,
and which is the more important  depends on
which induces the greater curvature $\kappa$ to the sound rays.
As Stokes and Reynolds realised  $\kappa  \approx \frac{1}{n} \frac{\p n}{ \p z} = \half 
\frac{1}{T} \frac{\p T} {\p z}$ for the thermal effect and 
$ \kappa  \approx \frac{1}{v_s}   \frac{\p v} { \p z}$ for the wind
shear, where $n$ is the index of refraction.  
It is probably no coincidence that Stokes's work followed
shortly after the first demonstration, by the German physicist 
Karl Friedrich Julius Sondhauss (1815 - 1886)     
in 1853 \cite{Sondhauss},  that a balloon filled with $CO_2$ will
act as a sound lens, focussing the sound of  a ticking watch 
so as to render  it audible some distance away.
 
\begin{figure*}
\begin{minipage}[t]{0.48\linewidth}
\centering {\includegraphics[width=2.5in]{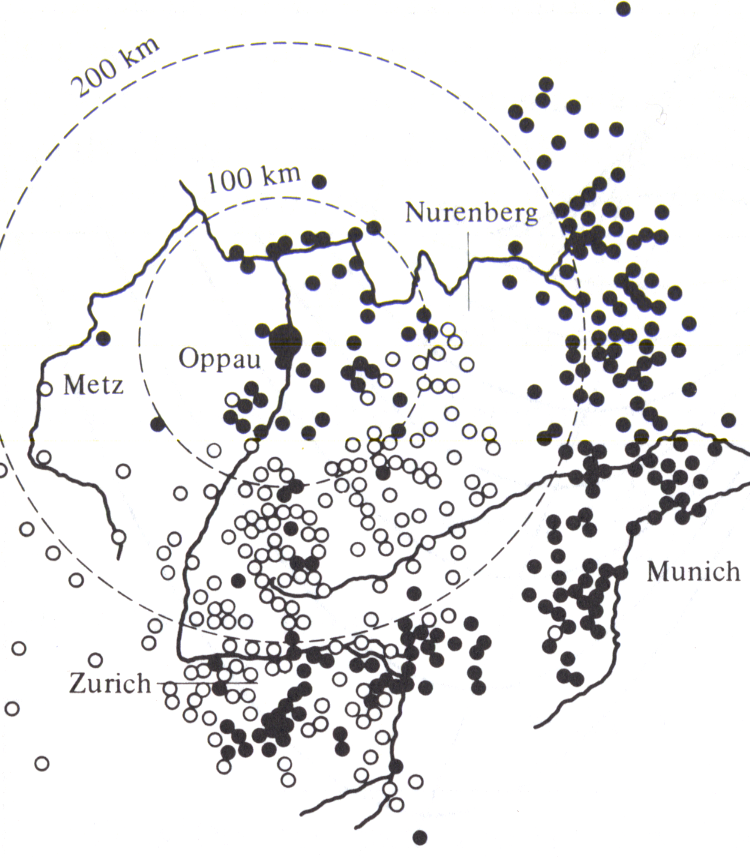}} \caption{The
  explosion at Oppau, 1921. Filled circles represent locations where
  the explosion was heard and empty circles locations where nothing
  was heard \label{oppaufig}}
\end{minipage}
\hspace{.3cm}
\begin{minipage}[t]{0.48\linewidth}
\centering 
\begin{picture}(0,0)
\put(-90,30){\includegraphics[width=2.5in]{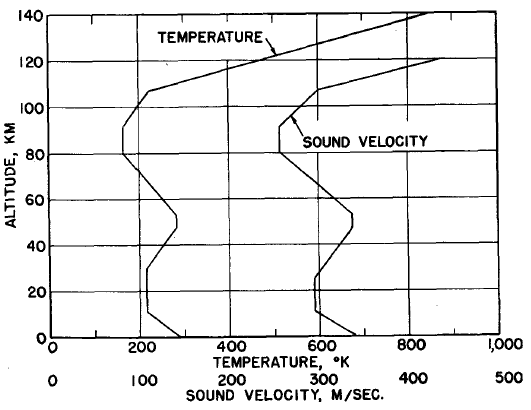}}
\end{picture}
\caption{The variation of temperature and sound speed with altitude\label{profilefig}}
\end{minipage}

\vspace{0.2cm}
{\emph{\small{Source: `Strange sounds in the Atmosphere: Part I',
  R.~K.~Cook, Sound 1, 12-l6 (1962).}}}
\end{figure*}

Examples of the interplay of these two effects, which can cause sound to travel
over large distances,  abound.
In early June, 1666, during the war between the Dutch and the English,  
both Samuel Pepys and John Evelyn reported in their diaries that
while the sound of gun fire from ships off the coast of Kent
could be heard clearly in London, it was not audible at all
at the ports of  Deal or Dover. As Pepys at the time observed:
\begin{quote}
``This \ldots makes room for a great dispute in Philosophy: how we
should hear it and not they, the same wind that brought it to us being
the same that should bring it to them.''
\end{quote}
Infrasound (sound of very low frequency) from the 
volcanic explosion of Krakatoa on August 27  1883 was heard to travel
 several times around the earth. During the first world war, the noise of the very large guns
on the Western Front  was often audible within a range of a 100  km
or so, and often beyond 200 km, but not within a 
 ``zone of silence'' between 100 and 200 km. On September 21, 1921 
there occured an enormous  explosion at Oppau  on the Rhine
and the same phenomenon was observed, see Figure \ref{oppaufig}.   

One  explanation for some of these phenomena is reflection of sound
from a layer of air in the upper atmosphere with a higher sound
velocity. The dependence of the velocity of sound in the  atmosphere  
follows its temperature profile. In the absence of wind, the
temperature (and hence velocity of sound) decreases
 under normal circumstances up to a height of about 10 km,
remains constant up to roughly 25 km, then incresases
to around 50 km, where it has a local maximum, and then has a 
local minimum at about 80 km, see figure \ref{profilefig}. A simple application
of the law of refraction
\ben
n(z) \sin \theta (z) = {\rm constant} \,, \label{refraction}  
\een    
where the local speed of sound $v_s(z)\propto \frac{1}{n(z)} $, 
reveals that sound rays can bounce off the maxima of $v_s(z)$ but can also be
trapped at the  minima. Presumably the latter effect was responsible for
the long distance propagation of infrasound from Krakatoa. A similar
phenomena occurs for sound waves in the ocean. The speed of sound
initially decreases with depth and then increases  exhibiting
a minimum value at a depth of around 1 km at the so-called 
SOFAR channel. This allows whales to communicate over very large distances
\cite{Morawetz, Weston}  and more sinisterly, submarines to snoop on other submarines.

For simple profiles $n(z)$, the law of refraction (\ref{refraction}) 
allows
simple solutions for the rays. Thus if $n(z) \propto z$,
one has catenaries with the horizontal axis $z=0$ as the directrix.
This gives a rough description of mirages in the desert. If
$n(z) \propto \frac{1}{\sqrt{z}}$ , the rays are cycloids, the curve
traced out by the point on the circumference of a circle rolling
without slipping along the horizontal axis (see Figure \ref{cyclfig}). One can imagine this as
the path of a glow-worm sitting on the rim of a bicycle wheel as it
rolls along in the dark. This could decsribe the rays passing over icebergs in the arctic.
More importantly for what follows later, this can also be achieved 
by assuming that $n(z) \propto \frac{1}{z}$, in which case the rays
are semi-circles  centred  on the horizontal axis. The addition of the
effects of wind complicates considerably this simple picture.

Considering for a moment fundamental physics, one of the clearest
trends in research over the
last 100 years or so has been what one might call the
\emph{geometrisation} of physics. This is by no means a new
phenomenon. Plato's association of the regular, or Platonic, solids with the
classical elements is perhaps the earliest attempt to explain the
world through geometrical intuition. Geometry, in the modern sense of
differentiable manifolds, really caught hold in physics in 1915 with Einstein's
general theory of relativity. Since then, the use of geometry has accelerated. Modern string theory, with
its $10$ space-time dimensions and complicated internal $6$-dimensional
Calabi-Yau manifolds is perhaps the clearest example of this
geometrisation process. 

A geometrical approach has much to recommend
itself, even to describe the more concrete physics we have thus far
been considering. In this article, we will discuss how the properties of
sound and light rays can be considered in a geometrical light. We will be
interested in the behaviour of solutions of the wave
equation\footnote{Throughout this paper we will use the convention
  that indices $i, j= 1 \ldots n$ which are repeated should be summed over} in a
moving medium: 
\ben
\Bigl[ \Bigl( \frac{\p}{\p t }  - W^i \frac{\p }{\p x^i}  \Bigr )^2 - 
c^2 h^{ij} \frac{\p ^2 }{\p x ^i \p x^j }  \Bigr ] u(x,t) =0\,. 
\een
Here $h^{ij}$ is related to the speed of sound in a direction parallel
to the unit vector $m_i$ by $v_{\mathbf{m}}=c(h^{ij}m_im_j)^{-1/2}$
and $W^i$ is a vector giving the velocity of the medium at each
point. We will be interested in particular in disturbances with short
wavelength, which move along \emph{rays}. We will start by considering
the case of a static medium, $W^i=0$. The geometry of the rays in this
case is well known and we shall describe some of its properties and
give some examples. We will then move on to the case of a moving
medium and demonstrate that this can also be described geometrically, as we
showed in \cite{GW1}. We will have to loosen slightly our notion of a
geometry in order to do so but the type of geometry which arises, a
Finsler geometry, is very natural. In fact, this geometry first arose
in our studies of light rays near rotating black holes. The link
between the effects of gravity on light rays and refraction of light
and sound waves by media can be made explicit and is the basis for
much current work on optical and acoustic black holes. These analogue
models for black holes allow experimentation in a laboratory, which would
of course not be possible with real black holes. For this reason, a
thorough understanding not just of the mechanisms of refraction, but
the \emph{geometry} of refraction is of great relevance both for
terrestrial and celestial physics.

\section{Geodesics and Fermat's Principle}

\subsection{Optical and Acoustic  Geometries}
The   elementary theory  of mirrors and lenses
is largely concerned with tracing the paths of  {\it light rays}
on reflection ({\it Catoptrics})  or 
refraction ({\it Dioptrics}) at a surface. Although the Greeks were uncertain
whether light proceeds from the eye to the object ({\it emission theory})
or from the object to the
eye {\it intromission theory}, 
and  whether its passage is instantaneous or at a finite
speed, nevertheless 
Heron of Alexandria (c.10-70 AD) 
was able to  formulate the laws of reflection
in terms of a {\it Principle of Shortest Length} from object to eye via
the reflecting surface. The occurrence of {\it caustics} 
shows that there can be more than one path, not all of which
are necessarily the shortest and  more accurately
we refer to the principle {\it Principle of Stationary Length},
that is the length of each ray is merely stationary among all neighbouring paths. 

Despite important pioneering  work by Ibn al-Haytham or Alhazen (965-c.1040)
demolishing the emission theory and investigating refraction,
it took longer to unravel the  fundamental law of Dioptrics. It    
it was not until independent work by
 Abu Sa`d al-`Ala' ibn Sahl (c.940-c.1000), Thomas Harriot (c.1560-1621)  and  Willebrord Snel Van Royen (1580-1626) that the familiar {\it Law of Sines}
was established and Pierre Fermat (c 1601- 1665)   
formulated his  unified {\it Fermat's 
Principle of  Stationary Time}. The idea is that the slowness of the ray
inside a medium  is proportional to its refractive index $n$. 
Note that the finite speed of light was only demonstrated by 
by Ole Roemer (1644-1710) using the eclipse  of Jupiter's moon Io 
in 1676.  The relation of the refractive
index to the speed of light  remained controversial
until experiments by Armand Hippolyte Louis Fizeau (1819-1896),
 Fresnel Augustin-Jean Fresnel (1788-1827) and  
George Biddell Airy (1801-1892) 
in the nineteenth century  finally established
its velocity as $\frac{c}{n}$, where $c$ is its speed in vacuo.
According to the discredited {\it corpuscular theory}
the opposite relation holds. Both theories give the same rays
but the speed with which light follows the rays differs. 
It would be more accurate therefore to speak of
 {\it Fermat's 
Principle of  Stationary Optical Path  Length }, where the optical length
of a path $\gamma$  is given by
\ben
L= \int _\gamma n  (\x) |d \x|    \,,  
\een  
where we have allowed for the possibility that the refractive index may
depend  upon position, as for example it does in a vertical stratifed medium
such as we encounter discussing mirages. Fermat's principle becomes
the {\it Variational Principle}  
\ben
\delta L= \delta \int _\gamma n  (\x) |d \x|  = 0\,.
\een 
By the time Fermat introduced this principle, Christiaan Huygens (1629-1695)
had initiated his wave theory of light and  
derived Fermat's Principle from it. His derivation
makes it clear   that it is the optical length or optical distance
which enters into all interference effects, and its therefore
appropriate to say that all optical measurements 
measure  {\it Optical Geometry}. By the same token,
measurements using sound waves  may be said to measure
{\it Acoustic Geometry} and  measurements  using seismic waves,
as on the earth or more recently the moon \cite{Weber}  to measure
{\it Seismic Geometry}. 
 
It is clear that these  geometries will, in an inhomogeneous
medium for which the dependence of $n$ on $\x$  is non-trivial,  
differ considerably from {\it Euclidean Geometry}. 
The existence of such {\it non-Euclidean 
Geometries} was first realised
by pure  mathematicians in the early part of the nineteenth century
working on the foundations of geometry. For centuries, 
people had been attempting to derive, starting from the other aximons,
Euclid's fifth axiom: that
through any point not on a given line there is exactly one line
parallel to the first. This seemed to them so obvious that it was
``neccessarily'' true. Eventually they gave up, 
Johann Carl Friedrich Gauss (1777-1855)  privately
and J\'anos Bolyai (1802 -1860)  and 
Nikolai Ivanovich Lobachevsky (1792-1856)  publicly,  
showed  the  existence
of two other types of homogeneous and isotropic 
 {\it Congruence Geometries}, Spherical and Hyperbolic. 
The first is easy to grasp in two dimensions  since it is 
just the geometry of the standard sphere, $S^2$  with the stationary  paths
(or {\it geodesics}) being the great circles.
Navigators, either by sea or air,   have been using spherical
geometry since at least the time of Columbus.  
It is not too difficult
to imagine a sphere in one dimension higher and indeed if 
the refractive index were to vary as 
\ben
n= \frac{n_0}{(  1 + \frac{\x ^2 }{4 R^2}   ) } \label{fish} 
\een       
the optical metric would be precisely that of three dimensional
spherical space $S^3$ of radius $n_0R$. 
The resulting optical  device is known as Maxwell's Fish Eye Lens
since  all rays emanating from any point $\x_e$  are circles
which  reconverge 
onto the  antipodal point $\bar \x _e= \frac{\x_e}{|\x _e|^2 }  $. We can verify that the optical distance along a 
radial geodesic is given by:
\begin{equation}
\int_{0}^{\infty} \frac{n_0}{ 1 + \frac{r ^2 }{4 R^2}} dr = \pi n_0 R,
\end{equation}
which is finite. Because,  for large enough
$|\x|$,  the refractive index $n$  drops below unity,    
the construction of such a device would require
the manufacture of a suitable ``meta-material''.
A more practical device was invented by Rudolph Karl Luneburg 
(1903-1949) 
and has 
\begin{eqnarray}
n&=&\bigl(2 |\bx|^2 -1)  \bigr ) \,, \quad |\bx | \le 1\,;
\nonumber \\ n&=&1, \quad  |\bx | \ge 1 \,.
\end{eqnarray}
This will focus all rays  incident on it a fixed direction
to the point on the circumference in the opposite direction.

The radius of curvature of the Spherical Geometry  given by 
(\ref{fish}) is $n_oR$.  To obtain Hyperbolic Space $H^3$ ,
often called {\it Bolyai-Lobachevsky space} or just {\it Lobachevsky
Space}, one needs only to  let the radius of curvature 
become pure imaginary which leads to 
\ben
 n= \frac{ n_0}{(  1 -\frac{\x ^2 }{4 R^2}   ) } \label{lob} \,. 
\een 
The refractive index becomes infinite when $|\x| = 2 R$ which
should be thought of as the  boundary at `infinity' of hyperbolic space.
In fact  the reader may easily verify that the optical distance along a 
radial path from the origin $\x=0$ to the boundary at infinity
is, by contrast with the case of spherical space, infinite.   

Jules Henri Poincar\'e (1854-1912) gave a simple analogy 
which illuminates the roles played by the flat Euclidean metric geometry
for which $n=1$  and the curved non-Euclidean geometry
for which $n$ is given by (\ref{lob}) as follows.
Imagine a medium  whose temperature varies with radius
as $ \frac{1}{n} $  
occupying a ball of radius $|\x|=2R$, as measured 
by a  measuring rod  made from a substance such as Invar
whose length is independent of temperature. 
If measured by a different ruler made of a material which expands in
proportion to the temperature, then the rod will shrink to zero length
as it approaches the  the boundary which is at zero temperature. The boundary
 will thus seem to be infinitely far away as measured by the second measuring rod.
At the time that Poincar\'e wrote, people were still reeling
under the discovery that what seemed to have been well established since 
antiquity:  that Euclid's  Geometrical Axioms were not logical necessities.
There was thus  great 
interest among geometers and philosophers as what was the ``correct''
or ``real''  geometry of space. For Poincar\'e it all depended upon how you
measure it. He did however believe that one should always
be able to find a system of measurements in which Euclid's Geometry holds.  

Despite appearances both Spherical Geometry and Hyperbolic
Geometry are, like their simpler Euclidean counterparts, 
both isotropic and homogeneous. For this reason
they are candidates for the physical geometry  of
space, as measured for example by light rays in vacuo. 
Indeed according to Einstein's theory of General Relativity
just these three possibilities can arise in the theory of the
{\it Expanding Universe} proposed by 
Alexander Alexandrovich Friedman (1888 -1925) and 
Monsignor Georges Henri Joseph \'Edouard Lema\^itre (1894 - 1966). 
For many years cosmologists have been attempting
to decide which best fits the observed universe. 
Based on observations of the
Cosmic Microwave Background (CMB), and other data, 
the consensus now is that it is flat Euclidean geometry. 
 
Of course no realistic  medium is exactly homogeneous or isotropic,
and  in particular  the speed of light, or sound, may depend upon direction
as well as position. A familiar example is 
provided by bi-refringence in crystals  such as calcite 
for which the ordinary and extra ordinary ray have different refractive indices,
$n_o$ and $n_e$.
This more general situation may be taken 
into account using a more general geometry invented by 
Georg Friedrich Bernhard Riemann (1826 - 1866) 
 called {\it Riemannian Geometry} in which
the optical or acoustic path  is given by 
\ben
\delta L=  \delta \int _\gamma \sqrt{h_{ij}(x_k)  \frac{ d x^i}{dt} \frac{dx^j} {dt} } dt  =0\,,
\een
where $h_{ij}$ is, in three spatial dimensions,  a $3\times 3$ 
symmetric array 
called the {\it  metric tensor}. Indeed one frequently introduces
what is called a {\it line element} which is an expression 
for the infinitesimal form of a generalised Pythagoras's theorem:
the infinitesimal distance $ds$  between points  $x^i$ and $x^i + d x^i$       is given by
\ben
ds^2 = h_{ij}(x_k) dx^i dx ^j \,. \label{metric} 
\een
Thus, for example,  in the case of a uniaxial  bi-refringent medium  
with unit  field $\bn$ in the ordinary direction the   
{\it Joets-Ribotta} 
optical  metric is \cite{Joets}   
\ben
ds^2 = n_e^2 d \x ^2 + (n_0^2 -n_e^2 ) (\n .d \x ) ^2 \,.
\een

Riemann's ideas are fully  incorporated into
Einstein's Theory of General Relativity. Indeed, for a static spacetime
a simple form of Fermat's Principle holds which, for example,  allows
one  to discuss the optics of black holes in terms of an effective
refractive index (in so-called isotropic coordinates) 
\ben
n(\x)= \Bigl(1 + \frac{G M}{2 c^2 |\x|}  \Bigr ) ^6 \Bigl 
( 1- \frac{G M}{2 c^2 |\x| } \Bigr ) ^{-2}\,,
\een
where $M$ is the mass and  $G$  is Newton's constant. In these
coordinates, the  black hole event horizon is located at 
$|\x|= \half \frac{GM}{c^2} $ . Note that, because
the  refractive index becomes infinite there, 
the event horizon is at  infinite  optical distance.
A closer examination reveals that as one approaches the event horizon, the
optical geometry approximates more and more closely  that of
hyperbolic space near its boundary at infinity as described above. 
We have recently pointed out that this is a universal
feature of all static (``non-extreme'') event  horizons 
and used it as a  quantitative tool for  
discussing some of their  puzzling physics \cite{GW2}.   

\begin{figure*}
\begin{minipage}[t]{0.48\linewidth}
\centering
\includegraphics[width=2.5in]{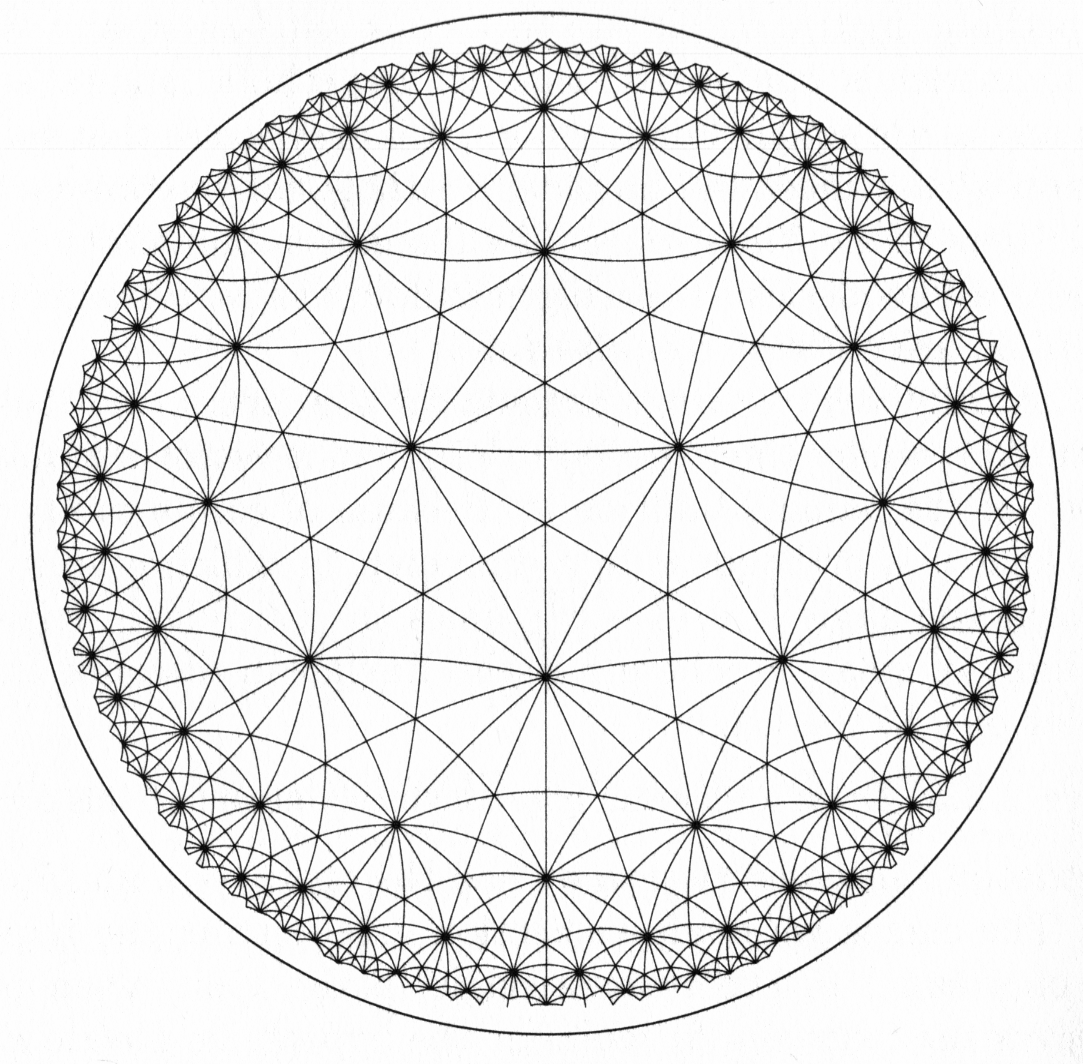} 
\caption{A  tiling of the Poincar\'e disk. The edges of the triangles are geodesics.  \label{H2tesball}}
\end{minipage}
\hspace{.3cm}
\begin{minipage}[t]{0.48\linewidth}
\centering
\begin{picture}(0,0)
\put(-90,30){\includegraphics[width=2.5in]{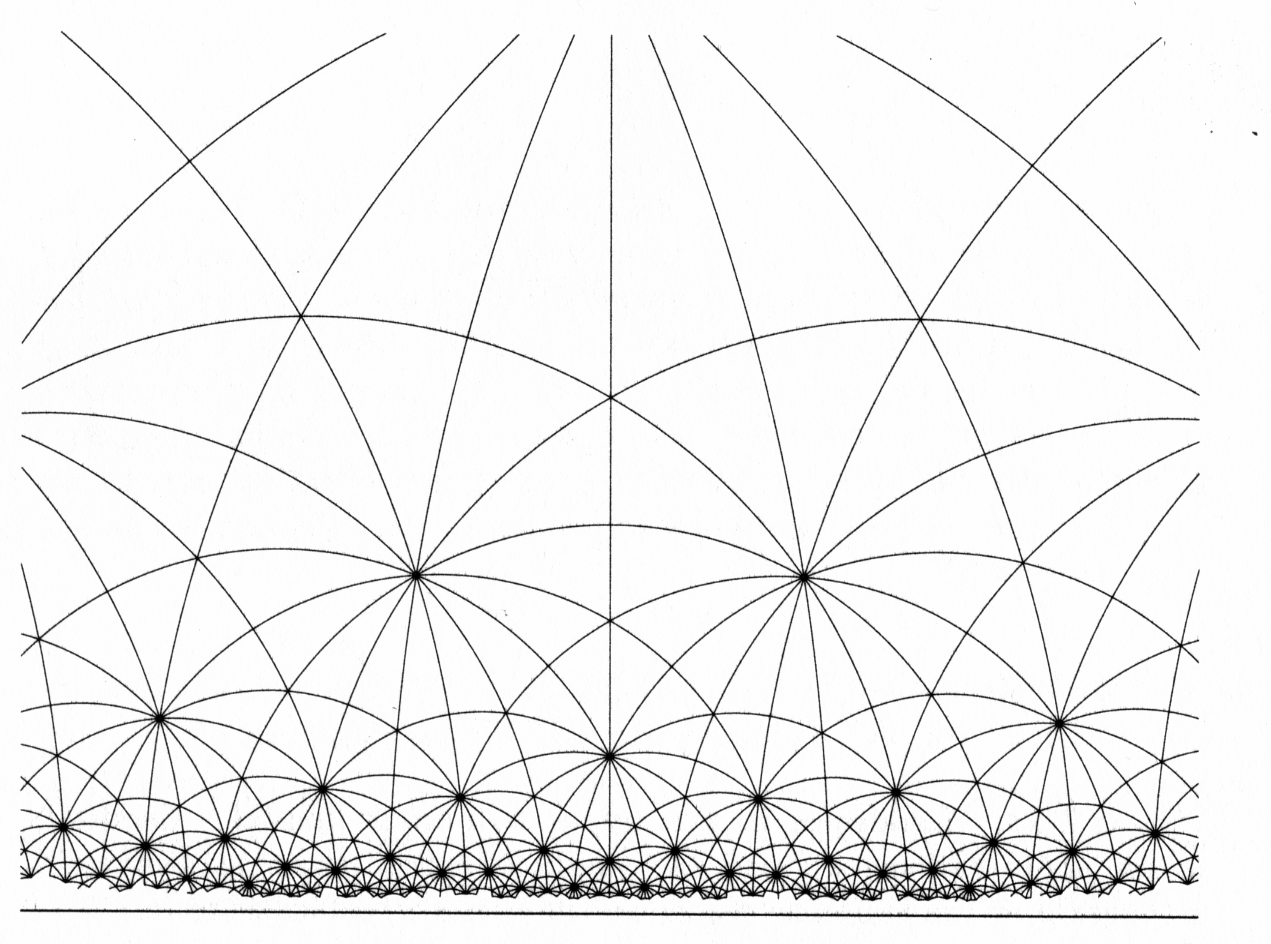}}
\end{picture}
\caption{The same tiling, this time shown on the upper half-plane
  model of hyperbolic space. \label{H2tesuhp}}
\end{minipage}

\vspace{0.2cm}
{\emph{\small{Source: Three Dimensional Geometry and Topology, Thurston}}}
\end{figure*}

Riemann also invented an even more general form of geometry,
taken up by Paul Finsler (1894-1970) called {\it Finsler Geometry} 
in which a more general expression
replaces (\ref{metric}) and this also arises in the optics and acoustics
of moving media as we shall discuss presently. In the meantime
we wish to say more about hyperbolic geometry, 
and in particular the hyperbolic plane $H^2$.

\subsection{The Hyperbolic Plane \label{hpsec}}

The geometry of the hyperbolic plane  is intimately connected
with that of the complex numbers. Their  introduction dramatically
simplifies many formulae. To see this we adopt units in which
the  radius of curvature $n_0R$ 
is set to unity.  We further set    
$\x= 2R  (x_1,x_2,x_3) $, $n_0=1$   in (\ref{lob}).
The {\it Hyperbolic Plane }, $H^2$, 
is obtained by setting $x_3=0$ . In order to exploit
the complex numbers we define  $z=x_1+ix_2$  to get 
\ben
ds ^2 = 4 \frac{ d x_1^2 + dx _2^2 } { 
(1 - x_1^2 - x_2 ^2 )^2 } = 4 \frac{ | d z |^2} {(1-|z|^2 )^2}\,.    
\label{disc} \een  
In this representation of the hyperbolic plane, the
 {\it Poincar\'e Disc}, $ H^2 $ occupies  the interior 
of the unit disc $|z|=1$  and one may verify that its geodesics 
are circular arcs which cut the  unit circle at right angles. A tiling of $H^2$ by triangles whose edges are geodesics is shown as Figure \ref{H2tesball}. The triangles are all similar, so one can see how the apparent length of a measuring rod shrinks as we approach the boundary.

\begin{figure*}
\centering
\begin{picture}(0,0)
\includegraphics{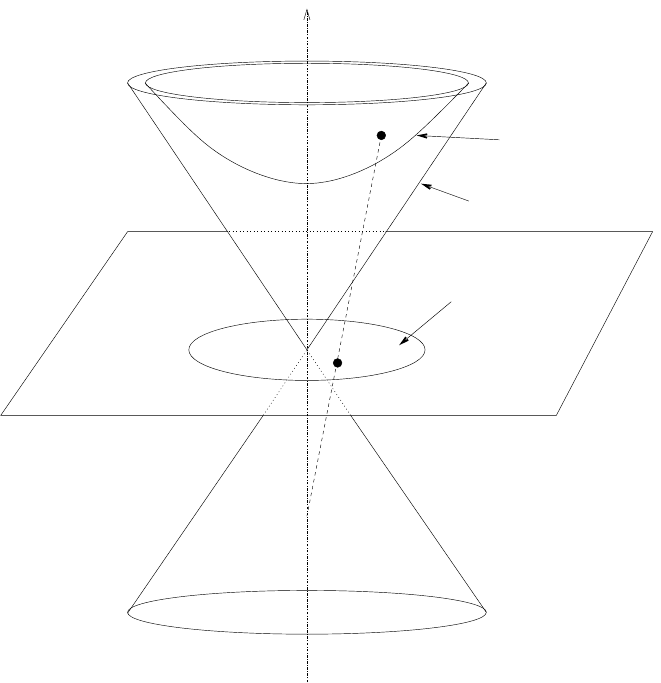}%
\end{picture}
\setlength{\unitlength}{1105sp}%
\begingroup\makeatletter\ifx\SetFigFont\undefined%
\gdef\SetFigFont#1#2#3#4#5{%
  \reset@font\fontsize{#1}{#2pt}%
  \fontfamily{#3}\fontseries{#4}\fontshape{#5}%
  \selectfont}%
\fi\endgroup%
\begin{picture}(11199,11670)(739,-12298)
\put(9376,-3061){\makebox(0,0)[lb]{\smash{{\SetFigFont{8}{6.0}{\rmdefault}{\mddefault}{\updefault}{\color[rgb]{0,0,0}Pseudo-sphere}%
}}}}
\put(8851,-4036){\makebox(0,0)[lb]{\smash{{\SetFigFont{8}{6.0}{\rmdefault}{\mddefault}{\updefault}{\color[rgb]{0,0,0}Light-cone}%
}}}}
\put(8626,-5836){\makebox(0,0)[lb]{\smash{{\SetFigFont{8}{6.0}{\rmdefault}{\mddefault}{\updefault}{\color[rgb]{0,0,0}Poincar\'e Disk}%
}}}}
\put(6151,-736){\makebox(0,0)[lb]{\smash{{\SetFigFont{8}{6.0}{\rmdefault}{\mddefault}{\updefault}{\color[rgb]{0,0,0}$X^0$}%
}}}}
\end{picture}%
\caption{The hyperbolic plane as the pseudo-sphere in Minkoswki
  space. Also shown are the light-cone $\mathbf{X} \cdot \mathbf{X}=0$
  and the plane $X^0=0$.\label{hypfig}}
\vspace{1cm}
\end{figure*}

We are free, in addition, to perform a coordinate transformation
to an equivalent  form. We choose a complex coordinate $w= u+iv$ 
related to $z$ by a {\it fractional linear transformation} 
\ben
w= i \frac{1+z}{1-z} \,.
\een   
This maps the unit disc into the {\it Poincar\'e Upper Half Plane}, $v >0$.
The centre of the unit disc maps to the point $w=i$ and the boundary
of the unit disc to the real axis $v = 0 $. In these coordinates the 
line element becomes 
\ben
ds ^2 = \frac{ |d w |^2 } { \Bigl | \frac{ w-\bar w} {2} \Bigl | ^2 } 
=   \frac{ du^2 + dv ^2 } { v^2 } \,. \label{upper} 
\een   
Because, as is easily verified, fractional linear transformations take 
circular arcs  to circular arcs, the geodesics are  again circular arcs. Because 
holomorphic maps are angle preserving, the geodeiscs are in fact  
semi-circles orthogonal to the real axis. Figure \ref{H2tesuhp} shows
the upper half plane tiled with the same geodesic triangles as in
Figure \ref{H2tesball}. 

Let's now think of $u$ as horizontal distance and
$v$  as vertical height in an  horizontally  stratified  medium
in which  the refractive  index $n$ decreases  with height.
If  we assume that over a certain range of heights, $v$, 
to a good approximation 
\ben
n \propto \frac{1}{v}  \label{law} 
\een
the rays will be semi circles. In this way we can easily explain
the mirage mentioned in the introduction, that in arctic or antarctic 
regions light is bent over 
icebergs and ships behind icebergs are observed to float 
upside in the air  above. To account for the mirages
seen in the desert or on hot days driving on motorways 
in which trees or cars seem to be reflected in  pools of 
still water,  it suffices to take the complex conjugate
and work in the  lower half plane.

Of course many laws of horizontal stratification $n=n(v)$
will give qualitatively similar results, but there is 
a considerable economy   to be made  by 
adopting the  law   (\ref{law}). If we do then the line element
(\ref{upper}) is invariant  under all fractional linear transformations
taking the upper half plane into itself. These are of the form 
\ben
w \rightarrow \frac{a w-b}{cw -d } \,, \qquad ad-bc=1,,  
\een 
where $a,b,c, d$ are real.  This defines the three dimensional group
$SO(2,1) \equiv SL(2,{\Bbb R}) /{\Bbb Z}_2$ 
isomorphic with the Lorentz Group of three dimensional  Minkowski spacetime. This is no accident. Much as it is convenient to think of the usual
$2$-sphere as the set of points at a fixed distance from the origin in
Euclidean $3$-space, there is a similar interpretation of hyperbolic
space as a \emph{pseudo-sphere} in Minkowski space. We consider the
space $\mathbb{R}^3$, but instead of the usual dot product, we endow
it with the Minkowski product:
\begin{equation}
\mathbf{X} \cdot \mathbf{Y} = -X^0 Y^0 + X^1 Y^1 + X^2 Y^2
\end{equation}
where $\mathbf{X} = (X^0, X^1, X^2)$ and similarly for $\mathbf{Y}$. For a
vector with $\mathbf{X}\cdot \mathbf{X}>0$, which we call
\emph{spacelike}, we can define the length to be $|\mathbf{X}| =
\sqrt{\mathbf{X}\cdot \mathbf{X}}$. The hyperbolic plane of radius $R$ is the set of
points defined by
\ben
\mathbf{X} \cdot \mathbf{X} = -R^2, \qquad X^0>0. \label{pseudosphere}
\een
The pseudo-sphere is sometimes referred to as the \emph{mass-shell}. This is because when we interpret the Minkowski spacetime as the geometry of special relativity in two spatial dimensions\footnote{In this case we work in units where the speed of light is $1$ and a particle of energy $E$ and momentum $(p_1, p_2)$ in the two spatial dimensions would have momentum vector $\mathbf{P} = (E, p_1, p_2)$.}, the vectors with $\mathbf{P} \cdot \mathbf{P}<0$ represent momentum vectors of particles whose rest mass, $m$, is given by $\mathbf{P} \cdot \mathbf{P}= -m^2$. The mass-shell thus represents all the possible velocities for a particle of a given rest mass. It is possible to check that any vector tangent to this surface is
spacelike, so that the Minkowski inner product allows us to define the
length of such a vector. The geometry of this surface with this
definition of length is that of the hyperbolic plane. A Lorentz
transformation leaves both the Minkowski metric and the condition
(\ref{pseudosphere}) unchanged and so represents an isometry of the
hyperbolic plane. To recover the Poincar\'e disk model, we
stereographically project the pseudosphere from the point $(-1,0,0)$
onto the plane $X^0=0$ as shown in Figure \ref{hypfig}.

\section{Zermelo's Problem and Finsler's geometry}

\subsection{Zermelo's navigation problem \label{Zermelosec}}

In the previous section, we considered the problem of finding the ray paths of a sound wave propagating in a static medium. We showed how Fermat's principle of stationary time leads us to consider the geodesics of a Riemannian metric. We would now like to consider how sound waves propagate through a moving medium. Fermat's principle continues to apply, however we must work a little harder in order to  translate the principle into a mathematical statement.

We will start off by considering a problem proposed by Ernst Zermelo
in 1931. Suppose a boat, which can sail at a constant rate relative to
the water on which it sits, wishes to navigate from point $A$ to point
$B$ as quickly as possible. If the water is at rest, then the captain
should steer along a straight line joining $A$ to $B$. More generally,
the captain should steer along a geodesic if the surface of the water
may not be taken to be flat, for example if the points A and B are far
enough apart that the Earth's curvature should be taken into
account. The navigation problem for the captain in this case
corresponds to finding the geodesics of some Riemannian metric.  

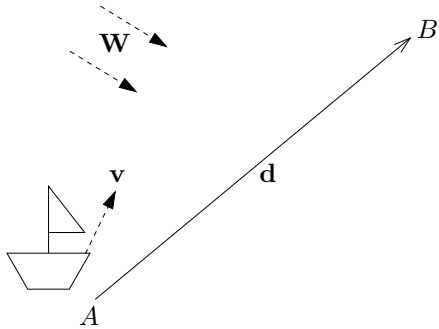
\begin{figure*}
\centering
\begin{picture}(0,0)%
\includegraphics{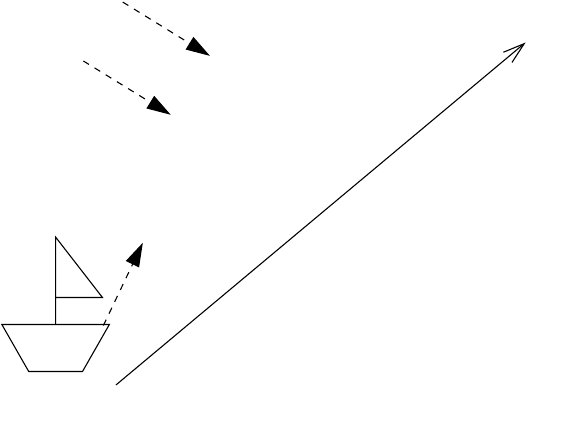}%
\end{picture}%
\setlength{\unitlength}{2763sp}%
\begingroup\makeatletter\ifx\SetFigFont\undefined%
\gdef\SetFigFont#1#2#3#4#5{%
  \reset@font\fontsize{#1}{#2pt}%
  \fontfamily{#3}\fontseries{#4}\fontshape{#5}%
  \selectfont}%
\fi\endgroup%
\begin{picture}(3986,2902)(3447,-4451)
\put(5701,-3136){\makebox(0,0)[lb]{\smash{{\SetFigFont{10}{9.6}{\rmdefault}{\mddefault}{\updefault}{\color[rgb]{0,0,0}$\mathbf{d}$}%
}}}}
\put(7100,-1850){\makebox(0,0)[lb]{\smash{{\SetFigFont{10}{9.6}{\rmdefault}{\mddefault}{\updefault}{\color[rgb]{0,0,0}$B$}%
}}}}
\put(4101,-4411){\makebox(0,0)[lb]{\smash{{\SetFigFont{10}{9.6}{\rmdefault}{\mddefault}{\updefault}{\color[rgb]{0,0,0}$A$}%
}}}}
\put(4276,-1970){\makebox(0,0)[lb]{\smash{{\SetFigFont{10}{9.6}{\rmdefault}{\mddefault}{\updefault}{\color[rgb]{0,0,0}$\mathbf{W}$}%
}}}}
\put(4368,-3137){\makebox(0,0)[lb]{\smash{{\SetFigFont{10}{9.6}{\rmdefault}{\mddefault}{\updefault}{\color[rgb]{0,0,0}$\mathbf{v}$}%
}}}}
\end{picture}%
\caption{Zermelo's navigation problem with a uniform drift\label{Zermdiag}}
\end{figure*}

Now let us suppose that the body of water on which the boat sits is
not at rest, but instead moves with some velocity
$\textbf{W}(\textbf{x})$, which we call the drift. The absolute
velocity of the boat is now the sum of two components: the motion of
the boat relative to the water, $\textbf{v}$, and $\textbf{W}$. In
order to find the fastest route between $A$ and $B$ the captain must
clearly take the drift into account. In order to do this, let us first
consider the simplest situation, where the surface of the water is a
plane and the drift $\textbf{W}$ is a constant vector, not changing
from point to point. This situation is shown in Figure
\ref{Zermdiag}. We will assume for convenience that the speed of the
boat relative to the water is $c$, a constant. We will
also assume that the speed of the drift is less than the speed of the
boat relative to the water, i.e. $\mathbf{W}^2 < c^2$. Let's define
$\mathbf{d}=\vec{AB}$ and work out how long it will take the boat to
get from $A$ to $B$ assuming that $\mathbf{v}$ is constant. In this
case, the position of the boat relative to $A$ at time $t$ will be 
\ben
\mathbf{x} = t (\mathbf{v}+\mathbf{W}).
\een
Supposing that the boat arrives at $B$ at time $T$, we have
\ben
\mathbf{d} = T (\mathbf{v}+\mathbf{W}). \label{dTeqn}
\een
We wish to solve for T as a function of $\mathbf{d}$ and
$\textbf{W}$. We can do this easily by looking at the equations we get
from dotting both sides of (\ref{dTeqn}) with $\textbf{W}$ and
$\mathbf{d}$: 
\bea
\mathbf{d}\cdot\mathbf{W} &=& T
(\mathbf{v}\cdot\mathbf{W}+\mathbf{W}^2), \\ \nonumber \mathbf{d}^2 &=&
T^2(\mathbf{v}^2+2 \mathbf{v}\cdot\mathbf{W} + \mathbf{W}^2). 
\eea
Noting that $\mathbf{v}^2=c^2$ and eliminating
$\mathbf{v}\cdot\mathbf{W}$ between the equations gives 
\ben
T[\mathbf{d}] =
\sqrt{\frac{\mathbf{d}^2}{c^2-\mathbf{W}^2}+\frac{(\mathbf{d}\cdot\mathbf{W})^2}{(c^2-\mathbf{W}^2)^2}}- 
\frac{\mathbf{d}\cdot\mathbf{W}}{c^2-\mathbf{W}^2}. 
\een
The condition that the speed of the drift is less than $c$ guarantees
that the denominators are positive and that $T[\mathbf{d}] \geq 0$,
with equality only when $\mathbf{d}$ vanishes. We notice that if
$\lambda>0$, then $T[\lambda\mathbf{d}]=\lambda T[\mathbf{d}]$. We can
also check that $T$ obeys a \emph{triangle inequality}: 
\ben
T[\mathbf{d}_1+\mathbf{d}_2] \leq T[\mathbf{d}_1]+T[\mathbf{d}_2].
\een
This means that the time $T$ which we have found is in fact the
\emph{least} time to travel between $A$ and $B$, because we cannot
reduce the time by travelling along the sides of a triangle with base
$AB$. A simple limiting argument shows that any curve from $A$ to $B$
will take longer to traverse than $T$.  

Now we can consider a more general problem, where the boat is
navigating in a Riemannian manifold $M$, with metric
$h=h_{ij}dx^idx^j$. The drift is a vector field $W^i(x)$ on $M$, which
may vary from point to point and whose length is always less than
$c$. Suppose the captain steers the boat so as to travel along a curve
$\gamma(s)$, with $\gamma(a)=A$, $\gamma(b)=B$. In order to find the
time taken to traverse this curve, we can approximate it with lots of
straight sections on which the metric and the drift are roughly
constant. We have worked out how long it takes to travel along such a
line segment and we can simply add these times up. Passing to a limit,
we find that the time taken to travel from $A$ to $B$ along this curve
is
\ben
T[\gamma] = \frac{1}{c}\int_a^b F[\gamma^i(s), \dot{\gamma}^i(s)] ds \label{time}
\een
where
\bea
F[x^i, y^i]/c &=& \sqrt{\alpha(x^i, y^i)}+\beta(x^i, y^i) \nonumber \\
\alpha &=& \frac{h_{ij} y^i y^j}{c^2-W^2} +\frac{(h_{ij}y^iW^j)^2}{(c^2-W^2)^2}\nonumber \\
\beta&=& - \frac{h_{ij}y^iW^j}{c^2-W^2}, \label{defF}
\eea
with $W^2 = h_{ij}W^i W^j$ This problem seems somewhat artificial as imagining a boat navigating
a general curved space is rather strange. This set up is very natural,
however, in the context of sound rays. If $h_{ij}$ is the acoustical
metric of a material (so that a ray moving in the direction of the
unit vector $n^i$ moves with velocity $c/\sqrt{h_{ij}n^i n^j}$) and
$W^i$ is a bulk motion of the material, then $T$ tells us the time the
sound would take to travel along $\gamma$.  We are now in a position
to make a mathematical statement of Fermat's principle for a sound
wave propagating through a moving medium. The sound rays are the paths
which extremise the time along the path, or equivalently the optical
length, $L=T c$:
\ben
\delta L = c \delta T = 0.
\een
Before we discuss what this means for the study of sound waves in a moving atmosphere, we will first discuss briefly a larger class of problems which have a similar form to ours. 

\subsection{Finsler and Randers geometry}

We have thus far been speaking somewhat loosely about geometries,
without describing exactly what we mean. For our
purposes, the particular type of geometry which is of interest is a
\emph{Finsler geometry}. Although named after Paul Finsler, the
concept of a Finsler geometry was introduced by Riemann in the same
lecture that he proposed what is now known as Riemannian geometry. The
defining feature of a Finsler geometry is that for suitably well
behaved curves, one can define a curve length. In order to do so, one
first has to define a  \emph{Finsler function}. The function $F$ of
(\ref{defF}) is a special case.  A Finsler function, in addition to an
assumption on its smoothness, is required to have three properties
\begin{itemize}
\item[1.] \emph{Positivity:} $F[x, y] \geq 0$, with equality only if $y=0$
\item[2.] \emph{Homogeneity:} $F[x, \lambda y] = \lambda F[x,y]$, for $\lambda \geq 0$
\item[3.] \emph{Subadditivity:} $F[x, y_1 + y_2] \leq F[x, y_1]+F[x, y_2]$
\end{itemize}
Given a Finsler function $F$, we can then define the length of a curve from $\gamma(a)$ to $\gamma(b)$ by:
\ben
L[\gamma] = \int_a^b F[\gamma^i(s), \dot{\gamma}^i(s)] ds. \label{finslen}
\een
Condition 1 ensures that the length of any non-trivial curve is
positive. Condition 2 ensures that the length of a curve does not
depend on its parameterisation. Condition 3 is necessary so that the
problem of finding curves of minimal length is well posed. This means
that we can talk about the \emph{geodesics} of $F$ as being curves
of minimal length between two points. Note that the `length' defined
in this way by the $F$ of the previous section is \emph{not} the usual
length of the curve, so the geometry defined by this $F$ differs from
the Euclidean geometry of the plane.

Whilst Riemannian geometry is fairly well understood, Finsler geometry
in general is much less well studied. This is mainly because of the
sheer variety of possible Finsler functions, which can be very
exotic. One of the simplest classes of Finsler function, into which
our function $F$ of the previous section falls, are the \emph{Randers
  metrics}. The Finsler function of a Randers metric is given in terms
of a Riemannian metric $a_{ij}$ and a one-form $b_i$ as
\begin{equation}
F[x,y] = \sqrt{a_{ij}(x) y^i y^j}+b_i(x) y^i. \label{Randers}
\end{equation}
This is a good Finsler function provided $a^{ij}b_i b_j<1$, where
$a^{ij}$ is the matrix inverse of $a_{ij}$.  One reason to study
Randers metrics becomes apparent when we consider the equations
satisfied by a curve which is a critical point of (\ref{finslen}) when
$a_{ij}$ is the flat metric in $\mathbb{R}^3$. Since we are free to
choose the parameterisation, we can assume that for the curve
$\mathbf{x}(s)$ we have $ \dot{\mathbf{x}}^2=1$ and we then find that
$\mathbf{x}(s)$ is a critical point when 
\ben
\ddot{\mathbf{x}} = \dot{\mathbf{x}}\times (\nabla \times
\mathbf{b}). \label{curleq}
\een
We see that $\mathbf{x}(s)$ follows the path of a particle of mass $m$
and charge $e$ moving in a magnetic field $\mathbf{B}= m/e(\nabla
\times \mathbf{b})$ with unit speed. For
this reason, the extreme curves of $L$ are sometimes referred to as
\emph{magnetic geodesics}. For a general $a_{ij}$, we find the
generalisation of the Lorentz force law in a curved space, with $b_i$
acting as the vector potential for the magnetic field. By extending
the notion of a metric to allow Finsler geometries, we have brought
the problem of charged particles in a magnetic field into the realm of
pure geometry. 

\subsection{A uniform magnetic field in the Hyperbolic Plane \label{unhyp}}

\begin{figure*}
\begin{center}
\includegraphics[width = .8\textwidth]{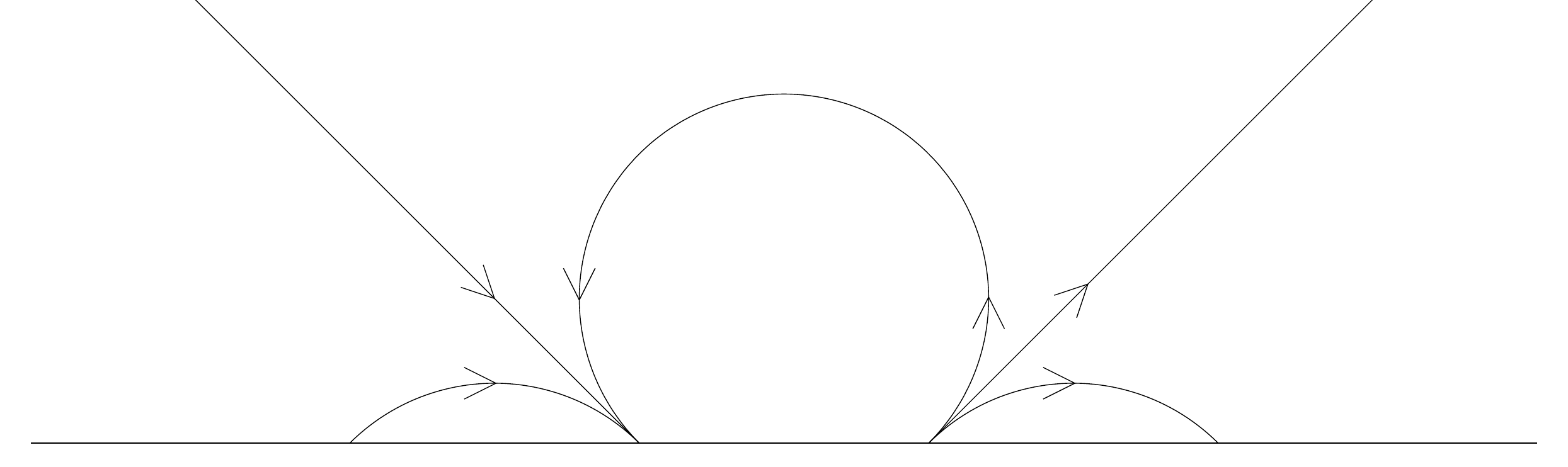}%
\end{center}
\caption{Some geodesics of the Randers metric given in (\ref{randmet1}), with $\alpha/\rho = 1/\sqrt{2}$.\label{magfig}}
\end{figure*}

As an interesting example, we take $x^1=u$, $x^2=v$, with $v>0$, and we will consider the Randers metric given by (\ref{Randers}) with:
\ben
a_{ij}= \frac{\rho^2}{v^2} \delta_{ij}, \qquad b_1 =  \frac{\alpha}{v}, \quad b_2 = 0. \label{randmet1}
\een
Where $\rho>0$ and $\alpha$ are fixed real numbers. This is a good Finsler metric, provided that $|\alpha|/\rho<1$. If $\alpha=0$, we see that this Randers metric in fact corresponds to the Riemannian metric considered in \S \ref{hpsec}, that of the hyperbolic plane of radius $\rho$ in its `upper half-plane' form. If $\alpha$ is non-zero, $b$ corresponds to a vector potential for the magnetic field which is everywhere directed straight out of the plane and which has strength $B=\alpha/\rho$, independent of position. Thus the geodesics of this Randers metric will correspond to a charged particle moving in a uniform magnetic field in the hyperbolic plane. In order to find the geodesics, we have to find curves $(u(s), v(s))$ which extremise the length
\ben
L = \rho \int ds \left(  \frac{\sqrt{\dot{u}^2+\dot{v}^2}}{v}+\frac{\alpha}{\rho} \frac{\dot{u}}{v}\right).
\een
Such a curve must satisfy the \emph{Euler-Lagrange} equations:
\begin{eqnarray}
0&=& \frac{d}{ds}\left( \frac{ \dot{v}}{v \sqrt{\dot{v}^2+\dot{u}^2}}\right) + \frac{\sqrt{\dot{v}^2+\dot{u}^2}}{v^2}+\frac{\alpha}{\rho} \frac{\dot{u}}{v^2}, \nonumber \\
0&=& \frac{d}{ds}\left( \frac{ \dot{u}}{v \sqrt{\dot{v}^2+\dot{u}^2}}+\frac{\alpha}{\rho }\frac{1}{v}\right). \label{ELeqns}
\end{eqnarray}
In the case where $\alpha=0$, we know from above that the geodesics are circles which meet the line $v=0$ at right angles. We also know that when we have a constant magnetic field in the usual Euclidean plane that the particle trajectories are circles. It seems reasonable then to guess that with $\alpha$ non-zero, the geodesics remain circular. We can consider a possible solution of the form
\begin{eqnarray}
u(s) &=& u_0 + r \cos s \nonumber \\
v(s) &=& v_0 \mp r \sin s.
\end{eqnarray}
Notice that the circle is traversed in a \emph{clockwise} or \emph{anti-clockwise direction} respectively for the two choices of sign. For a general Finsler metric, unlike for a Riemannian metric, the direction of travel is important and a curve will only be a geodesic when traversed in a particular direction.
Substituting into (\ref{ELeqns}) we find that we can satisfy the equations provided
\ben
\frac{v_0}{r} = \frac{\alpha}{\rho} = \pm B.
\een
Thus for $0<B<1$ we can either have clockwise circles with centres above $v=0$ or anti-clockwise circles with centres below $v=0$. Since $B<1$, we can interpret both cases geometrically as meaning that circles which meet the line $v=0$ at an angle $\theta = \cos^{-1} B$ are geodesics, provided they are traversed in the appropriate sense. Taking a limit where $r, v_0$ get larger and larger with their ratio fixed, we find that straight lines making an angle $\theta$ with the $v=0$ axis are also geodesics, provided again that they are traversed in the correct direction. A little more work shows that in fact any geodesic of this Randers metric is one of these curves. Figure \ref{magfig} shows examples of the various cases for $B = 1/\sqrt{2}$. For $B<0$, we can simply reverse the sense of the curves.

\section{Sound in a wind}

We noticed in \S \ref{Zermelosec} that Fermat's principle tells us that a sound ray in a medium with acoustic metric $h_{ij}$ with a wind $W^i$ will move along a geodesic of a related Randers metric defined by
\bea
&& a_{ij} = \frac{h_{ij}}{c^2-W^2}+\frac{W_i W_j}{(c^2-W^2)^2}, \nonumber \\ && b_i = -\frac{W_i}{c^2-W^2},
\eea
where $W^2 = h_{ij}W^i W^j$ and $W_i = h_{ij}W^j$.

Let's firstly consider what this means in the case where the speed of
sound is constant and equal everywhere to $c$ and the wind speed $W$ is small in magnitude compared to
$c$. This is a reasonable approximation for sound waves in a realistic
atmosphere. Typically $W/c < 0.03$ and even in the strongest
hurricanes, $W/c$ does not exceed $0.3$. For a constant, isotropic, speed of
sound, the acoustic metric is simply
\begin{equation}
h_{ij} = \delta_{ij}.
\end{equation}
If we work to first order in $w/c$, then we find that
\ben
a_{ij} = \frac{1}{c^2}\delta_{ij}, \qquad \mathbf{b} = - \frac{\mathbf{W}}{c^2}.
\een
Making use of (\ref{curleq}) and keeping track of the
factors of $c$, we find that the path followed by a sound ray is that
of a particle of mass $m$ and charge $e$ moving with speed $c$ in a
magnetic field given by
\begin{equation}
\mathbf{B} = -\frac{m}{e} (\nabla \times \mathbf{W}) = -\frac{m}{e}\bom,
\end{equation}
where $\omega$ is the vorticity of the wind. This justifies our
assertion in equation (\ref{vort}) that the vorticity of the wind acts
like a magnetic field on the sound rays.

A simple consequence of this correspondence is observed by
seismologists measuring the oscillations of the Earth after a large
earthquake. The spectral peaks are split by the Earth's rotation. From
our magnetic point of view, the effect of the rotation is to give rise
to a constant magnetic field inside the Earth. The splitting of the
spectrum is in precise analogy with the Zeeman effect which gives rise
to a splitting of spectral lines for atoms in a magnetic field.

\subsection{A stratified example}

\begin{figure*}
\begin{center}
\includegraphics[width=.8\textwidth]{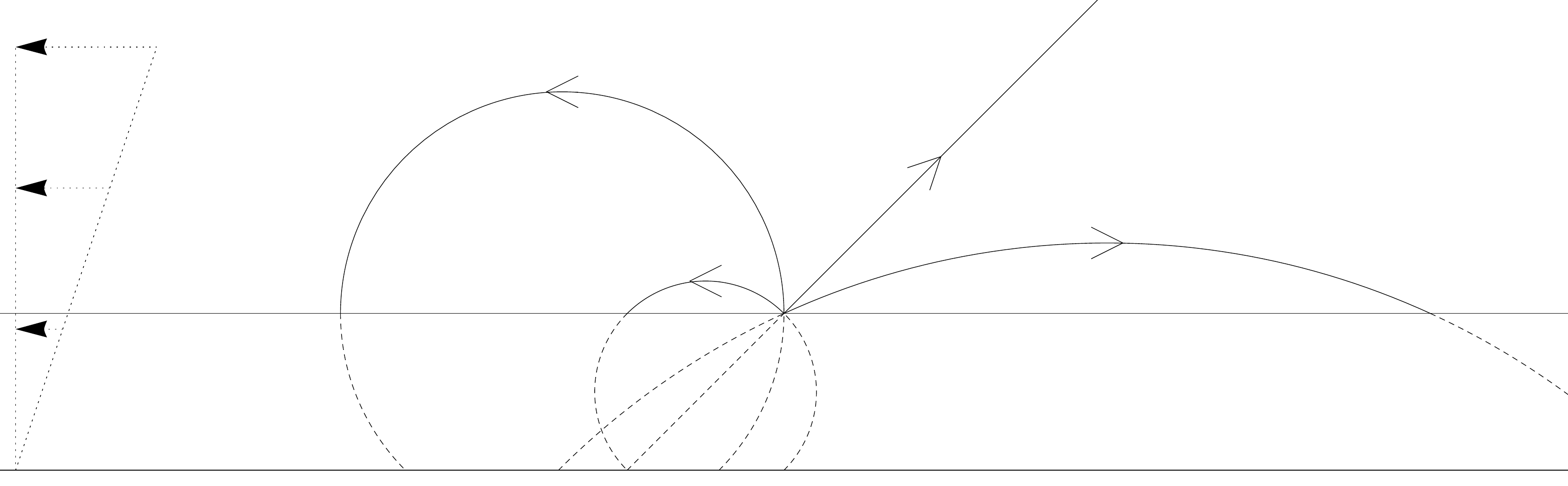}%
\end{center}
\caption{Some sound rays emanating from a point on the ground for $w/c = 1/\sqrt{2}, l=1$. The wind velocity at various heights (including the hypothetical extension below ground level) is shown at the left of the diagram. The rays are shown extending as dashed lines below ground level to show that they meet $z=0$ at an angle of $\pi/4$.\label{soundfig}}
\end{figure*}

An interesting problem to study involves combining a varying speed of
sound with a wind in a stratified atmosphere. We can gain some insight
by investigating a particular choice of acoustic metric and
wind. Let's suppose that the acoustic metric for the medium at rest
is the hyperbolic metric discussed previously
\ben
ds^2 = h_{ij}dx^i dx^j =l^2 \frac{dx^2 + dz^2}{z^2}
\een
We'll take this to model sound rays in an atmosphere varying with temperature near the ground, which we assume to be at $z=l$. The speed of sound at ground level is given by $c$. We'll suppose there's a horizontal wind with strength proportional to height, so that
\ben
\mathbf{W} = (-w \frac{z}{l}, 0).
\een
Where $w$ is some fixed parameter with units of velocity. We can easily calculate the associated Randers metric and find the time to go along a curve $(x(s), z(s))$. It is best expressed in terms of new variables:
\ben
 u= \frac{c x}{l\sqrt{c^2-w^2}} , \qquad v = \frac{z}{l}.
\een
This just represents a rescaling of the axes both parallel and perpendicular to the ground. In terms of $u, v$, the time is given by
\ben
T =  \frac{l}{\sqrt{c^2-w^2}} \int ds \left(  \frac{\sqrt{\dot{u}^2+\dot{v}^2}}{v}+\frac{w}{c} \frac{\dot{u}}{v}\right).
\een
Thus, the sound rays follow the geodesics of a Randers metric of precisely the form we considered in \S \ref{unhyp}, with  $B$ given by $w/c$!

As a simple application, we can consider the problem of tracing the
sound emanating from a point source, $P$ at ground level, which is at
$v=1$. We know that the sound rays in the $u, v$ coordinates are
circles (and lines) which meet the line $v=0$ at an angle $\cos^{-1}
(w/c)$. We also know that these circles have a clockwise sense if
their centre is above $v=0$ and an anti-clockwise sense if it's
below. This is already enough information to draw the rays shown in
Figure \ref{soundfig} for the particular value $w/c= 1/\sqrt{2}$. We
sketch outward directed geodesics which have a common starting point,
$P$. The straight line geodesic through $P$ plays an important role as
a \emph{separatrix}, which separates two different behaviours for the
trajectories. In this case, any ray emitted to the left of the
separatrix will return to the ground to the left of $P$, while a ray
emitted to the right returns to ground to the right. If we suppose
that $P$ is emitting sound uniformly in all directions, we see that
more of the energy emitted is absorbed by the ground on the left of
$P$ than on the right. In this case three times as much, which we can
see by considering the angle the straight geodesic makes with the
ground.

This example has three free parameters: $c$, $w$ and $l$. By choosing
these appropriately, we can match (for example) the speed of sound,
the speed of the wind and the the wind shear at ground level to a
real, more complicated profile. In fact, in \cite{GW1} we showed that
it is possible to construct a model based on the hyperbolic plane with
four free parameters, so that one can additionally set the rate of
change of speed of sound with height at ground level as well.

\section{Conclusion}

In this article we have sought to show that 
the motion of a charged particle moving in 
two-dimensional Lobachevsky space, or the hyperbolic plane,
equipped with a non-uniform  magnetic field can provide
a useful model for sound rays in a moving medium
with a gradient in the refractive index.
We have mentioned that in its three-dimensional version
the geodesics of Lobachevsky space provide   a useful model
for the motion of light rays near  the event horizon of a 
non-rotating black hole. If the black hole is rotating
then Coriolis type effects, referred to in General Relativity as the
rotation of inertial frames,   provide an effective magnetic field.
These two examples by no means exhaust  the possible applications
of hyperbolic geometry to physics. 

Two-dimensional surfaces abound in nature and if they have negative
Gauss curvature, sometimes  called {\it anti-clastic} at a point, the
surface cannot lie on one side of its tangent plane at that point.
Thus a finite smoothly embedded surface without edges  in Euclidean space
cannot have everywhere  negative Gauss curvature, but       
a finite portion of a surface with edges may. A simple example is 
provided by a
holly leaf. An example of great current  physical interest,
following the 2010  Nobel Prize to  Andre Geim and 
 Konstantin Novoselov is a graphene surface containing topological
defects called {\it disclinations} in
which some of the hexagonal lattice cells have been replaced by
heptagons. The electrical and other properties of such 
surfaces are  of great interest, and their study entails 
solving  the Dirac equation in a portion of 
two-dimensional Lobachevsky space.  The motion of charged particles
on  abstract finite  Riemann surfaces with no boundary or edges which  have 
constant  negative curvature and    uniform magnetic field 
are of interest in statistical mechanics since for week magnetic field
the motion is chaotic or {\it ergodic} as it is known technically.
However as the magnetic field strength is increased  there is a sudden
phase transition and this ceases to be the case. 

Three dimensional  Lobachevsky space has been invoked to model
some aspects of {\it quantum dots} and  the physics of 
four and five dimensional Lobachevsky space and their conformal
boundaries
are  currently of intense 
interest by String Theorists since Juan Maldacena suggested 
the famous {\it AdS/CFT correspondence} which has led to
a number of break throughs in quantum gravity and the 
quantum theory of black holes. Without going into 
technical details,  it may be of interest to outline 
some features of this fascinating idea.  Both in  Quantum Field
Theory and in String Theory it is customary to work in {\it imaginary
time} . Thus if we start 
in Minkowski spacetime with spacetime metric  
 \ben
ds ^2 = -c^2 dt ^2 + d{\bf x} ^2 \,, 
\een
we can pass to Euclidean space with positive definite metric
\ben
ds ^2 = c^2 d \tau ^2 + d {\bf x} ^2 \,.
\een   
 by setting
\ben
t=  i \tau \,, \qquad  \tau \, {\it real} \,.  
\een 
Often calculations may be performed more easily  
in Euclidean space. We then pass back to Minkowski spacetime  by setting 
 \ben
\tau =  - i t \,, \qquad  t \, {\it real} \,. 
\een
This process is called a {\it Wick Rotation} and it also works for
some curved spacetimes. A case in point is Anti-de-Sitter spacetime.
This is a solution of Einstein's equations with a negative cosmological
constant. It  may be obtained from Lobachevsky space by a simple Wick
Rotation. Maldacena's brilliant conjecture is  that there is 
a precise  correspondence  between  String Theory
in Anti-de-Sitter spacetime on the one hand,  and a special type
of quantum  field theory, called a Conformal Quantum Field Theory,
on the other hand,  the latter being  defined on the conformal boundary
of Anti-de-Sitter spacetime. The conformal boundary of 
Anti-de-Sitter spactime is  conformally  related to
Minkowski
spacetime   If we ``Wick rotate'' this conjecture
we are led to conjecture  a correspondence between  String Theory  in
Lobachevsky space and Quantum Field theory on its conformal boundary,
the latter being  conformally related to Euclidean space.

We hope that in this article  we have made it clear    
that not only is a knowledge of hyperbolic geometry and Lobachevsky  space
useful for understanding  trafffic noise,  but it has a much much wider 
range of applications in theoretical physics;  from cosmology to 
condensed matter physics to String Theory and  
Planck scale physics. We commend to the interested reader 
its study and further exploitation.

\end{multicols}

\section*{The Authors}

\begin{minipage}[t]{0.48\linewidth}
\begin{figurehere}
\begin{center}
\includegraphics{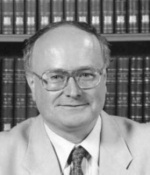}
\end{center}
\end{figurehere}
Gary Gibbons was born exactly 300 years after Gottfried Wilhelm
Leibniz. He came up to St. Catharine's College, Cambridge, to read
Natural Sciences in 1965, specializing in Theoretical Physics. After
taking Part III of the Mathematical Tripos he commenced research in
DAMTP, first under D.~W.~Sciama and then S.~W.~Hawking. After various
post-doctoral appointments he was appointed to a lectureship in DAMTP
in 1980. He is now Professor of Theoretical Physics in DAMTP. He was
elected to the Royal Society in 1999. He has been a Professorial
Fellow of Trinity College since 2002.
\vspace{1cm}
\end{minipage}
\hfill
\begin{minipage}[t]{0.48\linewidth}
\begin{figurehere}
\begin{center}
\includegraphics[height=1.75in]{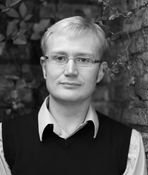}
\end{center}
\end{figurehere}
Claude Warnick came up to Queens' College, Cambridge, in 2001 to read
Mathematics. After completing Part III of the mathematical Tripos in
2005, he studied for a PhD in the General Relativity group of DAMTP,
under Gary Gibbons. He has been a Research Fellow at Queens' since 2008.
\end{minipage}

\end{document}